\documentclass[aps,eqsecnum,amsmath,onecolumn,notitlepage,nofootinbib]{revtex4}
\usepackage{graphics,setspace,epsfig,color}
\usepackage[letterpaper, dvips,width=7.5in,height=8.5in,includemp=false]{geometry}
\usepackage[vcentermath]{}
\usepackage{epsf}
\usepackage{amscd}

\newcommand{\beq}{\begin{equation}}
\newcommand{\eeq}{\end{equation}}
\newcommand{\bea}{\begin{eqnarray}}
\newcommand{\eea}{\end{eqnarray}}
\newcommand{\nn}{\nonumber}

\newcommand{\bigpar}[1]{\left(   #1 \right)}
\newcommand{\apara}{a_\parallel}
\newcommand{\ppara}{p_\parallel}
\newcommand{\aperp}{a_\perp}
\newcommand{\nb}{\mathbf{n}}
\newcommand{\mb}{\mathbf{m}}
\newcommand{\ab}{\mathbf{a}}
\newcommand{\pb}{\mathbf{p}}

\newcommand{\xb}{\mathbf{x}}
\newcommand{\yb}{\mathbf{y}}
\newcommand{\zb}{\mathbf{z}}
\newcommand{\xib}{\mathbf{\xi}}
\newcommand{\Deltab}{\mathbf{\Delta}}

\usepackage[normalem]{ulem}  % \sout{old text} for strikeout

\begin{document}

\title[title]{Friedel crystals and the outer crust of magnetars}
\author{ Paulo F. Bedaque }
\author{Simin Mahmoodifar}
\author{Srimoyee Sen}
\affiliation{Department of Physics \\
University of Maryland\\College Park, MD 20742}

%\date{}%
%\dedicatory{}%
%\commby{}%
% ----------------------------------------------------------------

\begin{abstract}
The strong magnetic fields found on the surface of magnetars are known to have profound effects in the physics of atoms in magnetar envelopes. We argue that the Friedel oscillations in the Coulomb force between the ions due to electron shielding can, for certain values of the parameters,  be the dominant effect determining the crystal structure in the outer crust of magnetars. We estimate the densities and magnetic fields for which this occurs, compute some of the elastic moduli and lattice phonon dispersion relations in this ``Friedel crystal" phase.
\end{abstract}

\maketitle

\section{Introduction}
The study of the structure and elastic properties of neutron star crusts is important in different astrophysical problems, in particular neutron star seismology. In magnetars, which are highly magnetized neutron stars with magnetic fields as large as $10^{15}$G on their surface, these elastic properties determine the properties of global seismic vibrations of their solid crust which are believed to be related to the observed quasi-periodic oscillations (QPOs) in the tails of giant flares \cite{Israel:2005av, Strohmayer:2005ks, Watts:2005ue} . 
The period of different oscillation modes of neutron stars  such as toroidal t-modes, spheroidal s-modes and interfacial i-modes as well as the coupling between the oscillations of the core and the crust of neutron stars  are very sensitive to the crustal properties \cite{ 1991ApJ...375..679S}. For example the period of the torsional oscillations of neutron star crust which can explain most QPOs, core-crust slippage which is important in estimating the Ekman layer damping rate of core r-mode oscillations \cite{Bildsten:1999zn,2006PhRvD..74d4040G}, as well as the corrections to the frequencies of the core oscillations due to the presence of a solid crust, all depend on the shear modulus of the crust.

In the relevant density regime for the outer crust of neutron stars, matter is composed of ions immersed in a sea of electrons (in the inner crust a gas of unbound neutrons is also present). At low enough temperature these ions arrange themselves into lattices whose structure determines many of the mechanical properties of the crust. Because of the importance of the elastic constants of neutron star crusts, there have been many computations of these quantities in the past. For example Fuchs \cite{1936RSPSA.153..622F} calculated the shear modulus of a static body centered cubic (bcc) Coulomb lattice, Ogata $\&$ Ichimaru \cite{1990PhRvA..42.4867O} calculated directionally averaged effective shear modus $\mu_{eff}$, which includes all elastic coefficients related to different shear and bulk moduli, for a bcc crystal at zero temperature using Monte Carlo  simulations and Strohmayer et. al. \cite{1991ApJ...375..679S} studied the temperature dependence of $\mu_{eff}$ in a classical one component plasma. Later Horowitz $\&$ Hughto \cite{Horowitz:2008xr} calculated the shear modulus of Coulomb crystals using molecular dynamics simulations, and taking into account electron screening in the Thomas-Fermi model. More recently Baiko \cite{Baiko:2011cb} computed $\mu$ using thermodynamic perturbation theory and taking into account ions motion. So far none of these studies have considered the effect of high magnetic fields on the lattice structure of ions and elastic properties of the neutron star crust. Here we concentrate on a particular effect of high magnetic fields on the elastic properties of magnetars crust.

In the outer crust of magnetars, the magnetic field strongly quantizes the motion of electrons perpendicular to the field into Landau orbitals. At low densities electrons occupy only the lowest Landau level. 
Sharma $\&$ Reddy \cite{Sharma:2010bx} studied the screening of ion-ion potential in a large magnetic field by calculating one-loop electron-hole polarization function.
They found that at low densities and high magnetic fields, similar to the outer crust of magnetars, the screening length for Coulomb interactions between ions can be smaller than inter-ion spacing. More importantly they found that the screening is anisotropic and the ion-ion potential along the magnetic field has a long-range oscillatory behavior (Friedel oscillations). 
Even though  the Friedel oscillations are weak they are much longer ranged than the screened Coulomb potential and, consequently can have important consequences for the crystal structure, as conjectured in \cite{Sharma:2010bx}. It is the purpose of this paper to explore this possibility and some of its consequences.
%This long-range potential along the magnetic field will change the structure and elastic properties of the crystal in the outer crust of magnetars. 

%In the absence of magnetic fields ions in the outer crust of neutron stars  are assumed to form a bcc lattice. At large magnetic fields (B $> 10^{14}$ G) and low densities ($\rho < 10^{8} g/cm^3$) when only the lowest Landau levels are filled, the long-range Friedel oscillations can change the crystal structure to something unusual which will be discussed in a separate paper. The equation of state of magnetized Coulomb plasmas has been recently computed in Ref.~\cite{2013A&A...550A..43P}, but the effect of magnetic field on the ion-electron (electron polarization) contribution to the free energy of the plasma haven't been included in those computations.
We will make a simple, heuristic but quantitative argument arguing that, for a large region of the parameter space, the Friedel oscillations make the ions  organize themselves into strongly coupled filaments parallel to the magnetic field with spacing along the magnetic field being $a_\parallel=\pi/k_e$ ($k_e$ is the electron Fermi momentum)\cite{Sharma:2010bx}. These filaments interact with each other more weakly and the crystal structure in the direction transverse to the magnetic field is more uncertain.
We compute some of the elastic constants in the outer crust of magnetars that are dominated by the longitudinal structure of the lattice and can be calculated without knowing the exact crystal structure. 
%Since only the direction parallel to the magnetic field is important in our computations, here we consider a simple tetragonal lattice to compute elastic constants.

The paper is organized as follows. In section I we discuss ion-ion potential, in particular the electronic screening of ions in the presence of large magnetic fields and crystal structure at densities relevant to the outer crust of magnetars. In section II we discuss elastic properties of magnetar crust and compute two elastic constants that are dominated by the longitudinal structure of the lattice. In section III we compute the dispersion relation of lattice phonons moving in the direction parallel to the magnetic field. Finally we provide a summary of our findings and conclusions in section IV.

\section{Friedel oscillations and crystal structure at high magnetic fields}

In the density regime of interest ($10^6 {\rm g/cm}^3 \alt \rho \alt 10^8 {\rm g/cm}^3$) the distance between ions is much smaller than atomic orbits, the atoms are fully ionized and the system is composed of ions surrounded by an electron gas. The expectation is that at small enough temperatures the repulsion between the ions, modified by the intervening electrons, will make the ions settle on a lattice. The properties of the electron gas, and consequently of the inter-ion potential, are strongly modified by magnetic fields. So we will start our discussion considering electrons in a magnetic field.

The single particle states of an electron in a magnetic field are labelled by two continuos momenta $p_y$ and $p_z$, a positive integer $n=1/2, 3/2, \cdots$ (Landau level) and a spin orientation $m_s$ ($m_s= 1/2$ for the lowest Landau level $n=1/2$ and $m_s=\pm 1/2$ for $n=3/2, \cdots$), with an energy equal to $\epsilon=\sqrt{p_z^2+m^2+2 e B n}$, where $m$ is the electron mass and $B$ the external magnetic field (we use units where $c=\hbar=k_B=1$). These states are centered in one of the directions transverse to the magnetic field at $y=p_y/eB$. The many-body ground state is formed by filling the  first Landau level with increasing values of $p_z$ until the energies reach the $p_z=0$ value of the {\it second} Landau level. From that point on the lowest and the second lowest Landau level are filled simultaneously. The number of single-particle states in a box of size $L^3$ is
\beq\label{eq:ne}
n_e = \frac{1}{L^3}\frac{L eB}{2\pi/L}\frac{L}{2\pi} \int_{-k_e}^{k_e} dp_z = \frac{eB}{2\pi^2}k_e,
\eeq where $k_e$ is the z-component of the momentum of the most energetic electron (the Fermi momentum in the direction of the magnetic field). From eq.~(\ref{eq:ne}) we can find the mass density $\rho$ at which the electrons become relativistic
\beq
 \rho \approx \frac{A M m eB}{2\pi^2 Z} \approx 0.79\times 10^8 {\rm g/cm}^3 \left(\frac{A_{66}}{Z_{28}} \right)B_{15},
\eeq where $B_{15}=B/10^{15}G$, $M$ is the nucleon mass  and we express the mass ($A$) and atomic ($Z$) number of the ions in terms of $A_{66}=A/66$  and $Z_{28}=Z/28$ (chosen because  $^{66}Ni_{28}$ is one of the favored isotopes at these densities).

We are interested in the regime where all (or most) electrons are in the lowest Landau level. This corresponds to mass densities below
\beq
\rho \alt \frac{A M}{\sqrt{2}\pi^2 Z}(eB)^{3/2} \approx  5.2 \times 10^8 {\rm g/cm}^3\ \left(   \frac{A_{66}}{Z_{28}}\right)B_{15}^{3/2}.
\eeq 
%Electron can also be excited to higher Landau levels by temperature effects. The separation between the bottom of the first and second Landau level is
%\beq
%\sqrt{m^2+eB}-\sqrt{m^2+3eB}
%\eeq
Thus, for almost all the parameter space we are interested in, most electrons are in the lowest Landau level and the effect of the magnetic field on electron motion is appreciable. 

The effect of the magnetic field on the electron screening of Coulomb forces was discussed in \cite{Sharma:2010bx}. The potential between ions was found to be
\beq\label{eq:V_ion}
V(r_\perp, z) = Z^2 \alpha
\left[
\frac{   e^{-m_D \sqrt{r_\perp^2+z^2}}  }{\sqrt{r_\perp^2+z^2}}
- \frac{m_D^2 e^{-z/\lambda_T} }{4 z} \cos(2 k_e z){\sqrt{4k_e^2+\frac{m_D^2}{2}\log(4k_e z)}}  r_\perp K_1\left(r_\perp \sqrt{4k_e^2+\frac{m_D^2}{2}\log(4k_e z)}\right)
\right],
\eeq where $m_D^2=e^3Bm/(2\pi^2 k_e)$ and $\lambda_T = 2\pi k_e/(mT)$.  The direction along the magnetic field is denoted by $z$ and the perpendicular direction  is $r_\perp$. Eq.~\ref{eq:V_ion} is valid for $z\gg 1/k_e$ and $r_\perp \gg 1/\sqrt{ e B}$. We will require an expression valid for the $r_\perp \ll 1/\sqrt{ e B}$ regime as well. The value of $V(r_\perp=0,z)$ can be found by going back to  a form more generally valid \cite{Sharma:2010bx}. The Yukawa part remains the same while the Friedel part is given by:
\bea
V_F(0, z) &=& 
-\frac{Z^2 \alpha m_D^2}{2}  \frac{e^{-z/\lambda_T}  \cos(2 k_e z)}{z}
\int_0^\infty dq_\perp    \frac{     J_0(q_\perp r_\perp)    e^{-\frac{q^2}{2eB}}    }
{\left(q_\perp^2+4 k_e^2+\frac{m_D^2 e^{-\frac{q^2}{2eB}}}{2} \log(4k_ez)\right)^2+\left(\frac{\pi m_D^2}{2}e^{-\frac{q^2}{2eB}} \right)^2}\nn\\
&\approx&
 -\frac{Z^2 \alpha m_D^2}{4}  \frac{e^{-z/\lambda_T}  \cos(2 k_e z)}{z}
 \frac{1}{4 k_e^2+\frac{m_D^2}{2}  \log(4k_ez)}
 f\left(\frac{2k_e^2+\frac{m_D^2}{4}\log(4k_e z)}{eB}\right),
\eea with
\beq
f(x) = 1+x e^x E_i(-x),\label{eq:f}
\eeq   $E_i(x)=\int^\infty_x e^{-t}dt/t$ being the Exponential Integral function. The function $f(x)$ approaches $1$ as $x\rightarrow 0$ and $1/x$ as $x\rightarrow\infty$.

Two important points are worth noticing about eq.~{\ref{eq:V_ion}}. The first is that the potential  can be decomposed into an isotropic part describing a shielded Coulomb potential (denoted by Yukawa part from now on) and an
 anisotropic  one describing the Friedel oscillations. The second point is that eq.~\ref{eq:V_ion} contains several length scales. The parameter $\lambda_T$ is by far the longest one. Numerically it is given by
 \beq
 \lambda_T = \frac{2\pi k_e}{mT} \approx 1.6\times 10^{-7} cm \left(  \frac{Z_{28}}{A_{66}}    \right) \bigpar{\frac{\rho_8}{B_{15} T_1}},
 \eeq where $T_1=T/keV$ and $\rho_8=\frac{\rho}{10^8 g/cm^3}$ , while the other scales are
 \bea
 \frac{1}{m_D}  &\approx &  1.3\times 10^{-10}\ cm\  \sqrt{\frac{   Z_{28} \rho_8}   {A_{66}  }  }\frac{1}{B_{15}},\\
 \frac{1}{k_e}   &\approx &  3.0\times 10^{-11}\ cm\  \bigpar{  \frac{A_{66}}{Z_{28}}    } \bigpar{  \frac{B_{15}}{\rho_8}    },\\
 \frac{1}{\sqrt{eB}}  &\approx & 8.1 \times 10^{-12}\ cm \ \frac{1}{\sqrt{B_{15}}}.
 \eea This disparity in length scales means that the Friedel term is very long ranged compared to either the screening length $1/m_D$ or the inter-ion distance $n^{-1/3}$. As such, it can be added coherently over a large number of ions and dominate over the larger but short-ranged Yukawa term in determining the crystal structure. If that is the case,  the ions are spaced by the distance $a_\parallel = \pi/k_e$ in the longitudinal direction in order to be at the bottom of the $\cos(2 k_e z)$ oscillations (a scenario already suggested in \cite{Sharma:2010bx}) . The lattice structure in the transverse direction is less clear but also less important for what follows. This picture of the crystal structure will be the central assumption of our work.  Monte Carlo studies are essential in verifying that the potential in eq.~{\ref{eq:V_ion}} indeed leads to such a lattice structure  but they are somewhat involved. Here we will present a plausibility argument supporting our assumption.  
 
 %%%%%%%%  parametersapce %%%%%%%%%%%%%%%%%%%%%%%%%%%
\bigskip
\begin{figure}[t]
  \centerline{\includegraphics[width=80mm]{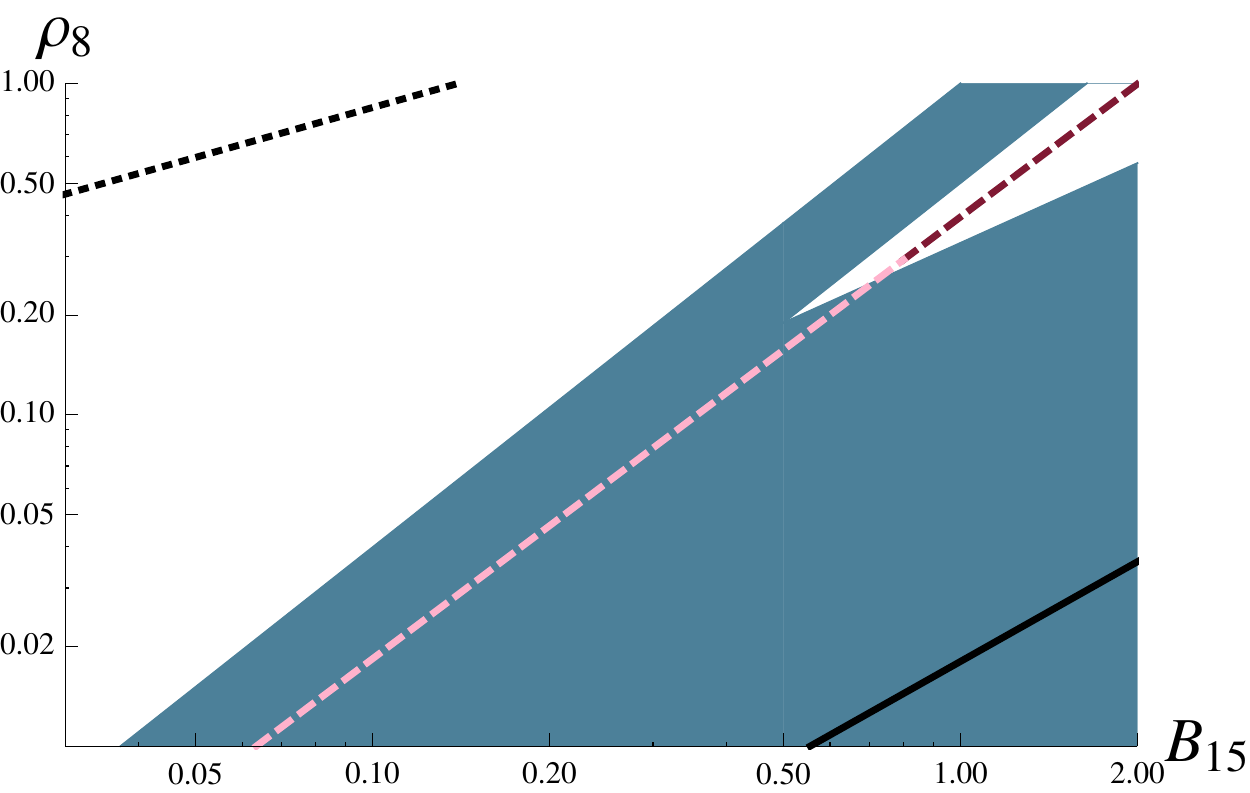}}
%\vskip 0.15in
\noindent
\caption{ \label{fig:parameterspace}
The darker region (blue online) is the approximate regions in  $\rho_8 \times B_{15}$ space where our simple model predicts a ``Friedel crystal". Below the dashed line the approximation $m_D^2\log(4k_e \lambda_T) \ll 8k_e^2$ is valid.
 We used $Z_{28}=A_{66}=T_1=1$ (there is a mild logarithmic dependence on the temperature on this graph). In the region above the dotted line on the upper left corner and below the solid line in the lower right corner, function $f$ defined in eq.~\ref{eq:f} can not be approximated by 1. The white triangular region on the upper right corner corresponds to a situation where the minimum of the free energy of the crystal happens at $\apara \neq \frac{\pi}{k_e}$, and therefore ``Friedel crystal'' doesn't exist anymore. This has been illustrated in Fig.~\ref{fig:Eion}.
 }
\end{figure}  
%%%%%%%%%%%%%%%%%%%%%%%%%%%%%%%%%%%%%%%%%%%%%

First, let us observe that the expression in eq.~{\ref{eq:V_ion}} can be simplified by noting that
 \beq
 \sqrt{4k_e^2+\frac{m_D^2}{2}\log(4k_e z)} \approx 2 k_e
 \eeq for all $z$ such that $z\ll \lambda_T$, that is, for all $z$ such that the Friedel term is not negligible. In fact, at $z\approx \lambda_T$ we have
 \beq
\frac{\frac{m_D^2}{2}\log(4 k_e \lambda_T)}{4 k_e^2} \approx 0.006 \bigpar{  \frac{A_{66}^3 B_{15}^4}{Z_{28}^3  \rho_8^3} 
\log\bigpar{\frac{2.1 \times 10^4 Z_{28}^2 \rho_8^2}{A_{66}^2B_{15}^2 T_1}},}
 \eeq 
 therefore eq.~{\ref{eq:V_ion}} can be approximated by
 \beq\label{eq:V_ion_ap}
V(r_\perp, z) = Z^2 \alpha
\left[
\frac{   e^{-m_D \sqrt{r_\perp^2+z^2}}  }{\sqrt{r_\perp^2+z^2}}
- \frac{m_D^2 }{8  k_e} \frac{\cos(2 k_e z) e^{-z/\lambda_T} }{z} r_\perp K_1\left(2k_e r_\perp\right)
\right].
\eeq in the region of the parameter space where $B_{15}^4 \alt 10 \rho_8^3$ (the region below the dashed line in fig.~\ref{fig:parameterspace}) and $r_\perp \gg 1/\sqrt{eB}$. At $r_\perp=0$ this formula is modified to
\beq
V(0, z) = Z^2 \alpha
\left[
\frac{   e^{-m_D z}  }{z}
- \frac{m_D^2 }{16  k_e^2} \frac{\cos(2 k_e z) e^{-z/\lambda_T} }{z}  f\left(\frac{2k_e^2}{eB}\right)\right].
\eeq

Let us consider a tetragonal ion lattice (a cubic lattice stretched/contracted in the $z$ direction) with the spacing longitudinal and perpendicular to the magnetic field equal to $a_\parallel $ and $a_\perp $ respectively, and related to each other in such a way to keep the ion density $n=1/(a_\perp^2 a_\parallel)$ fixed and equal to $n_e/Z$). Both terms in eq.~{\ref{eq:V_ion}} decay fast in the $r_\perp$ direction, but the Friedel part decays very slowly in the longitudinal direction.  Thus the potential energy of one ion in the lattice 
can be approximated by the Yukawa contribution from its nearest neighbors  and the sum of the Friedel part along the z-axis:
\beq\label{eq:E_ion}
E_{ion} = 4 V_{Yukawa}(z=0, r_\perp=a_\perp) +2 V_{Yukawa}(z=a_\parallel,r_\perp=0)+ 2\sum_{n=1}^\infty 
V_{Friedel} (z=n a_\parallel, r_\perp=0),
\eeq  The Friedel contribution is cutoff by the factor $e^{-z/\lambda_T}$ thus the largest relevant value of $z$ is $\sim \lambda_T$. At those values the arguments of the function $f$ is
\beq\label{eq:condition}
\frac{2 k_e^2+\frac{m_D^2}{4}\log(4 k_e \lambda_T)}{e B} \log(4 k_e \lambda_T)
\approx
0.14 \bigpar{   \frac{   Z_{28}^2 \rho_8^2}     {   A_{66}^2  B_{15}^3    }}+  0.001 \bigpar{     \frac{A_{66} B_{15}}   {Z_{28} \rho_8        }} \log\left(  \frac{2.1 \times 10^4 Z_{28}^2\rho_8^2}{A_{66}^2 B_{15}^2 T_1} \right).
\eeq The argument of $f$ is then small for most of the parameter space except for the $B_{15}\ll 1$, $\rho_8\sim 1$ and a small region where $B_{15} \agt 50 \rho_8$. Thus $f\alt 1$ in the regions above the dotted line or below the solid lines in fig.~\ref{fig:parameterspace}. As we will see now, these regions are not important for our purposes as a Friedel crystal is not likely to be formed there. In those regions the Friedel force is somewhat diminished as compared to the approximation leading to eq.\ref{eq:E_ion} but the Friedel crystal may still exist there.

\begin{figure}[t]
\begin{minipage}[t]{0.45\linewidth}
\centering
\includegraphics[width=\textwidth]{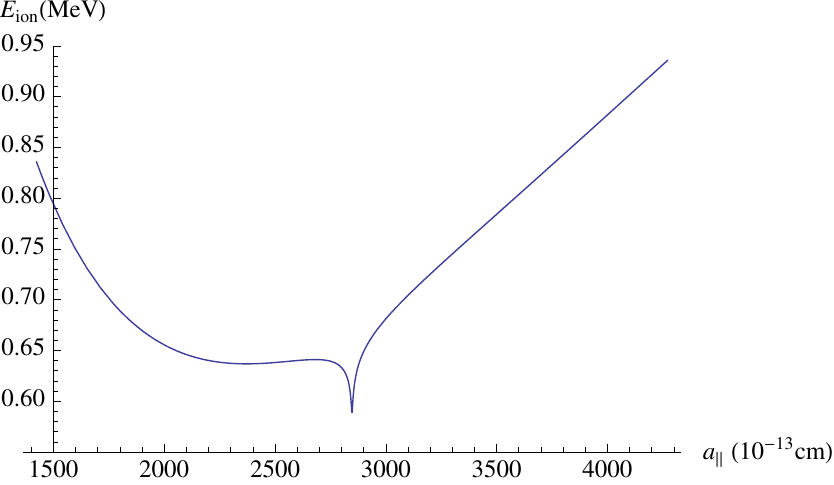}
\label{fig:figure1}
\end{minipage}
\hspace{0.5cm}
\begin{minipage}[b]{0.45\linewidth}
\centering
\includegraphics[width=\textwidth]{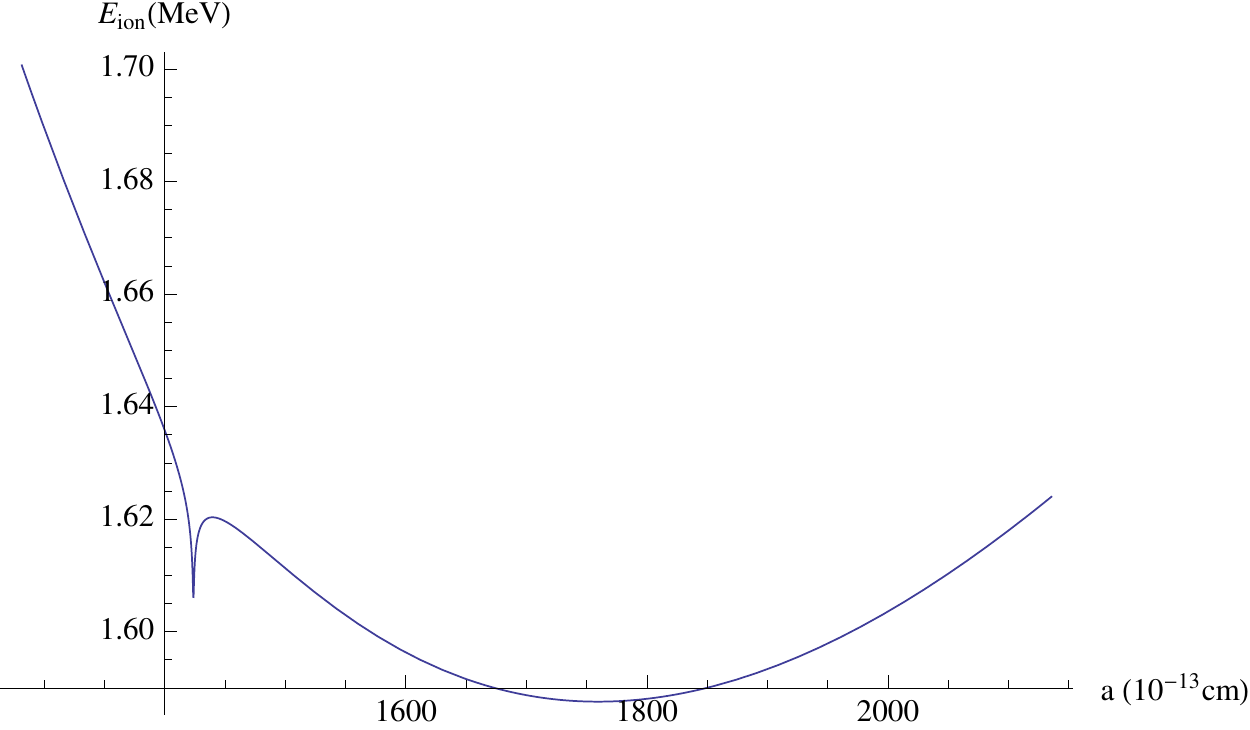}
\label{fig:figure2}
\end{minipage}
\caption{\label{fig:Eion}
Energy per ion as a function of $a_\parallel$ for $B=3\times 10^{14} G$, $\rho_8=10^7 g/cm^3$ (left) and $\rho_8=2\times10^7 g/cm^3$ (right). The remaining parameters are $T=1\ keV, A_{66}=1, Z_{28}=1$. The global minimum of $E_{ion}$ happens at $\apara=\frac{\pi}{k_e}$ in the left plot, but it happens at $\apara \neq \frac{\pi}{k_e}$ in the right plot. Therefore there is no ``Friedel crystal'' for the parameters used in the right plot.
}
\end{figure}

The sum  in eq.~\ref{eq:E_ion} is readily evaluated with the result
 \bea
E_{ion} &=&
Z^2 \alpha 
\left[
4 \sqrt{na_\parallel} e^{-\frac{m_D}{\sqrt{na_\parallel}}} + 2 \frac{ e^{-m_Da_\parallel} }{a_\parallel}
-\frac{m_D^2}{16 k_e^2 a_\parallel} 
\left(
\frac{a_\parallel}{\lambda_T} - \log2-\log(\cosh\frac{a_\parallel}{\lambda_T} -\cos(2k_e a_\parallel))
\right)
\right]
\eea where in the last line we used the fact that $a_\parallel \ll \lambda_T$. This expression breaks down at $a_\parallel \ll 1/k_e$. At those short distances there is no screening and the inter-ion interaction is described by a simple repulsive Coulomb potential. The  behavior of this expression as a function of $a_\parallel$ is shown, for two representative combinations of the parameters, in fig.~\ref{fig:Eion}. 
For low enough temperatures, the logarithmic singularity at $\apara=\pi/k_e, 2\pi/k_e, \cdots$  (arising from the sum over the long-range potential proportional to $1/z$ and cut off by the thermal screening at the scale $\lambda_T$) guarantees a minimum of $E_{ion}$ \footnote{At small temperatures one can minimize $E_{ion}$ instead of minimizing the free energy to get the crystal structure.} at these values of $\apara$, which is more pronounced at $\apara=\pi/k_e$. For temperatures relevant to magnetar crusts this minimum may not be a global minimum or it may disappear all together. The issue is the competition between the weaker but longer-ranged Friedel part of the potential with the Yukawa part. In order to keep the density fixed and $\apara=\pi/k_e$, the transverse lattice spacing $\aperp=\sqrt{\frac{2 \pi Z}{e B}}$ is independent of density. As the density increases (at fixed magnetic field), $\apara$ decreases and when it is comparable to the range $\sim 1/m_D$ of the Yukawa potential it becomes energetically favorable for $\aperp$ to increase and $\apara$ to decrease. For larger values of the magnetic field, above $B\approx 5\times 10^{15}$ G, there is even a more complicated phase structure with the ``Friedel crystal"  disappearing and appearing again as the density changes due to competition between the Friedel potential and the Yukawa interaction both in the longitudinal and transverse directions. 

We estimate the regions of parameter space favorable to the formation of the ``Friedel crystal" by (globally) minimizing eq.~\ref{eq:E_ion}. The result is shown in fig.~\ref{fig:parameterspace}. We stress that the arguments above are only heuristic. The spacing by $\apara=\pi/k_e$ along the longitudinal direction is likely to survive a full analysis as it depends only on the fact that the potential decays fast on the perpendicular direction while it has long-range Friedel oscillations in the longitudinal one. But the transverse structure of the lattice, taken here to be a square, is considered  purely for illustration purposes and may well turn out to be more complicated after a further analysis. Also,  the precise locations of the phase boundaries may change upon a more detailed analysis. In particular, the physics determining the transverse and longitudinal structures of the lattice are different and there may be  temperature range where the transverse structure melts while the more rigid longitudinal structure remains. In this regime the crust would be a nematic liquid crystal, ordered in the direction of the magnetic field but disordered in the transverse direction. We are presently studying these possibilities
\footnote{In the absence of a magnetic field the Friedel oscillations in the potential have the form $\sim \cos(2 k_e r)/r^3$, where $r$ is the radial distance. The higher power of $r$ compared to the denominator of eq.~\ref{eq:V_ion} is compensated by the fact that the oscillations occur in all directions and a {\it potential} infrared divergence in the energy   $\sim \int dr r^2 \cos(2 k_e r)/r^3$ could be important in determining the crystal structure. However, in three dimensions it is impossible to arrange the ions so that their mutual distance is always an integer multiple of $\pi/k_e$, so the contributions from the Friedel part can not act coherently and the Friedel part of the potential is of limited importance. Only in the effectively one-dimensional case discussed in the  paper can we arrange the ions so that their mutual distance is always a multiple of $\pi/k_e$ and the Friedel part of the potential can add up to a dominate factor.}. In the regions of the parameter space where $a_\parallel \ll \lambda_T$ does not hold, the effect of the Friedel oscillations do not add coherently over many ions and we do not expect the crystal structure to be valid. In that case, a standard lattice determined by the repulsive (screened) Coulomb force is more likely. Notice that the regions where our simple model predicts the existence of a Friedel crystal the condition that eq.~\ref{eq:condition} is much smaller than one is satisfied.

%Consequently, the potential well around the minimum at $\apara=\pi/k_e$ becomes deeper as the temperature is lowered
%\beq
%E_{ion}(\apara=\frac{\pi}{k_e}) \approx -Z^2 \alpha \frac{m_D^2}{16\pi k_e} \log\left(  \frac{2 k_e^2}{mT}\right).
%\eeq

%%%%%%%%  Eion   %%%%%%%%%%%%%%%%%%%%%%%%%%%
\bigskip

%%%%%%%%%%%%%%%%%%%%%%%%%%%%%%%%%%%%%%%%%%%%%

At finite temperature the ions fluctuate around their equilibrium positions. When the fluctuations are of the order of the lattice spacing the lattice order is broken and the lattice melts. We can estimate the temperature in which this occurs as following. When an ion is displaced along the longitudinal direction by an amount $x$ its potential energy changes to
\bea
E_{ion}(x) &\approx&
 \sum_{n=1}^\infty V_{Friedel} (z=n \pi/k_e-x, r_\perp=0) + \sum_{n=1}^\infty V_{Friedel} (z=n \pi/k_e+x, r_\perp=0)\nn\\
&\approx&
 E_{ion}(x=0) + \frac{Z^2 \alpha m_D^2 k_e}{4\pi}  \log(\frac{k_e \lambda_T}{\pi}) \ x^2+\cdots
\eea where we kept only the interactions with other ions in the $z=0$ line, the dominating force fixing $\apara=\pi/k_e$. Equating the average potential energy $E_{ion}(x)$ to the temperature ($E_{ion}(x)\sim T$) we can find the typical thermal displacement of the ion and imposing that this displacement is smaller than $\apara$ we find
\beq
T_{melt} \approx \frac{Z^2 \alpha m_D^2}{k_e}  \log(k_e\lambda_T)|_{T=T_{melt}}\approx 
\frac{Z^2 \alpha m_D^2}{k_e}  \log(\frac{2k_e^2}{mT_{melt}})|_{T=T_{melt}}
\approx 1\  MeV   \bigpar{\frac{A_{66}^2 B_{15}^3}{\rho_8^2}},
\eeq far above the temperatures of interest. Quantum fluctuations can also destabilize the lattice. To estimate this effect we compare the mean-square-root fluctuations of the ion displacement in the ground state to $\apara=\pi/k_e$. We find
\beq
\frac{\sqrt{\langle x^2\rangle}}{\apara} \approx \left(  \frac{k_e^3}{AMZ^2 \alpha m_D^2 \log(k_e \lambda_T)} \right)^{1/4}
\approx
0.08 \bigpar{\frac{Z_{28}^2 \rho_8^4}{A_{66}^5 B_{15}^5}}^{1/4},
\eeq that is small for all but the smallest magnetic fields and highest densities of interest.

 %%%%%%    ELASTIC PROPERTIES

\section{Elastic properties}

Elastic properties of magnetar crusts are important in the modeling of the oscillations of magnetars. In this section we compute some of the elastic constants -- the ones dominated by the longitudinal structure of the lattice -- in the crust lattice model discussed  in the previous section.

In a tetragonal lattice with lattice spacings $\apara$ and $\aperp$ the equilibrium position of the ions are given by $\ab.\nb = \aperp n_x \hat\xb + \aperp n_y \hat\yb+\apara n_z \hat\zb$, where $\hat\xb,\hat\yb$ and $\hat\zb$ form an orthonormal basis and $n_x, n_y$ and $n_z$ are integers. We denote by $\xib_\nb$ the displacement of the ion located at $\ab.\nb$ from its equilibrium position. For small displacements $\xi_\nb \ll \aperp,\apara$ the potential energy of the lattice is given by

\bea
U  &=& \frac{1}{2} \sum_{\nb, \mb} V((\nb-\mb).\ab + \xib_\nb - \xib_\mb)\nn\\
&\approx&
U_0 +  \frac{1}{4} \sum_{\nb, \mb} (\xi^i_\nb-\xi^i_\mb)(\xi^j_\nb-\xi^j_\mb) \
\underbrace{
\frac{\partial^2 V}{\partial r^i\partial r^j} |_{\xib=0}
}_{= V^{ij}_{\nb\mb}}
+\cdots ,
\eea 

where $U_0$ is the equilibrium energy. It is convenient to rewrite $V^{ij}_{\nb\mb}$ in terms of the difference $\nb-\mb$:

\beq
V^{ij}_{\nb\mb} = \sum_\Deltab V^{ij}_\Deltab \delta_{\nb, \mb-\Deltab   }
\eeq and we obtain

\bea
\Delta U = U-U_0 &=& \frac{1}{4} \sum_{\nb,\Deltab} V^{ij}_\Deltab (\xi^i_\nb- \xi^i_{\nb+\Deltab})  (\xi^j_\nb- \xi^j_{\nb+\Deltab}).
\eea Using the Fourier components

\beq\label{eq:fourier}
\xib_\pb = \aperp^2\apara \sum_\nb e^{-i \pb.\ab.\nb}  \xi_\nb
\qquad\text{and}\qquad
\xib_\nb = \int \frac{d^3p}{(2\pi)^3} e^{i \pb.\ab.\nb}  \xi_\pb,
\eeq where $\pb.\ab.\nb= (p_x n_x+p_yn_y)\aperp + p_z n_z \apara$ we find

\beq
\Delta U = \frac{1}{\aperp^2\apara }  
\underbrace{
\int_{-\pi/\aperp}^{\pi/\aperp}  \int_{-\pi/\aperp}^{\pi/\aperp}  \int_{-\pi/\apara}^{\pi/\apara} 
\frac{d^3p}{(2\pi)^3}
}_{=\int_B \frac{d^3p}{(2\pi)^3} }
 \sum_\Deltab V^{ij}_\Deltab   \sin^2\left(\frac{\pb.\Deltab.\ab }{2} \right)    \xib^i_\pb \xib^j_{-\pb} .
 \eeq As the lattice contracts or expands it carries the electron gas, to whom they are strongly coupled. Thus the elastic deformation energy will also receive a contribution from the electron gas compressibility. This contribution can be computed as
 
 \bea
 \epsilon(n_e(1+\nabla.\xib)) - \mu_e n_e &\approx&  \underbrace{\epsilon(n_e) - \mu_e n_e }_{{\rm constant}}
 + 
 \underbrace{
 n_e \left(  \frac{d\epsilon}{dn_e}-\mu_e \right)   \nabla.\xib
 }_{=0} + \frac{1}{2} (\nabla.\xib)^2 
 \underbrace{
 n_e \frac{d^2\epsilon}{dn_e^2}
 }_{=\kappa}+\cdots,
 \eea where $\epsilon(n_e)$ is the energy density of the electron gas and $\kappa$ its compressibility. Using the Fourier components in eq.~\ref{eq:fourier} we find that the energy change due to the electron gas is
 \beq
 \Delta U_e = \frac{\kappa}{2\aperp^2\apara} \int_Bd^3p \sum_{i,j=1,2,3}   \frac{\sin(p_ia_i)\sin(p_ja_j)}{a_ia_j}\xi^i_\pb  \xi^j_{-\pb}
 \eeq with $a_{1,2}=\aperp$ and $a_3=\apara$. For a free electron gas in a magnetic field we have
 \beq
 \kappa = \frac{k_e^2}{\sqrt{k_e^2 + e B + m^2}}
 \approx
 0.42\ MeV \bigpar{\frac{Z_{28}^2 \rho_8^2}{A_{66}^2 B_{15}^2}} \frac{1}{\sqrt{0.26 + 5.8 B_{15} + 0.42 \bigpar{\frac{Z_{28}^2 \rho_8^2}{A_{66}^2 B_{15}^2}}}}
 \eeq
  
  For deformations varying little on the lattice spacing scale we can expand the elastic energy in powers of momenta and find the total energy of
 the deformed lattice:
 
 \bea
 \Delta U + U_e &=&
  \frac{1}{\aperp^2\apara}\int_B \frac{d^3p}{(2\pi)^3} 
 \left[
 \sum_\Deltab V^{ij}_\Deltab \sin^2\left(  \frac{\pb.\Deltab.\ab}{2}\right) +\frac{\kappa}{2} \frac{\sin(p_ia_i)\sin(p_ja_j)}{a_ia_j}
 \right]
 \xib^i_\pb  \xib^j_{-\pb}\nn\\
     &\approx&
      \frac{1}{\aperp^2\apara}\int \frac{d^3p}{(2\pi)^3} 
 \left[
\frac{1}{4} \sum_\Deltab V^{ij}_\Deltab p_k p_l \Delta_k  \Delta_l a_k a_l+\frac{\kappa}{2} p_i p_j \right]
 \xib^i_\pb  \xib^j_{-\pb}\nn\\
      &\approx&
       \frac{1}{\aperp^2\apara}\int \frac{d^3p}{(2\pi)^3} 
 \left[
 \frac{1}{4}\sum_\Deltab V^{ij}_\Deltab \Delta_k \Delta_l a_k a_l+\frac{\kappa}{2} \delta^i_k \delta^j_l \right]
 p_k p_l  \xib^i_\pb  \xib^j_{-\pb}\nn\\
        &\approx&
\frac{1}{2}  \int d^3x\  \lambda^{ij}_{kl}\ \partial_k \xi^i \partial_l \xi^j,
 \eea 
 
 where 
 \beq\label{eq:lambda}
 \lambda^{ij}_{kl}=\frac{1}{\aperp^2\apara} \left(
 \kappa\ \delta^i_k \delta^j_l + \frac{1}{2} \sum_\Deltab V^{ij}_\Deltab
\Delta_k \Delta_l a_k a_l
 \right)
 \eeq and the constants $\lambda^{ij}_{kl}$ are the elastic moduli tensor as defined, for instance, in \cite{:fk}.
 The elastic moduli tensor encapsulates the information about the elastic properties of the lattice. In the case of an isotropic medium, they are determined by only two independent constants, bulk and shear moduli. In general, the lattice breaks rotation symmetry down to a discrete subgroup and  more independent constants are necessary to specify $\lambda^{ij}_{kl}$. In the case of a tetragonal lattice, for instance, there are 6 independent ones. In our case, beside the symmetry breaking due to the lattice there is the breaking of rotation symmetry due to the external magnetic field. Thus, the same lattice, oriented in space in a different direction relative to the field, will have a different energy. As such, the energy does not depend only on the symmetric combinations $(\partial_k \xi_i + \partial_i \xi_k)/2 $ forming the strain tensor, but also on the anti-symmetric combinations $(\partial_k \xi_i - \partial_i \xi_k)/2 $ describing rigid rotations of the lattice.

 %%%%%%%  lambda zzzz
 
Since we have not determined the crystal structure that is energetically most favorable, in this paper we will restrict ourselves to the computation of $\lambda^{zz}_{zz}$ and $\lambda^{zz}_{xx}$ only, since they, as we will argue, are largely independent of the transverse structure of the lattice. They will also determine the dispersion relation of phonons moving in the longitudinal direction and the heat conductivity along the field. $\lambda^{zz}_{zz}$ is the elastic constant related to the energy change of the lattice due to a small displacement in the z direction (parallel to the magnetic field) when the wave is also propagating in the z direction, and $\lambda^{zz}_{xx}$ is related to the energy change of the lattice when the displacement is in the x direction (perpendicular to the magnetic field) but the wave is propagating in the z direction.
We first compute $\lambda^{zz}_{zz}$ analytically  by making some approximations valid for the $B_{15}^4 \alt 10 \rho_8^3$ case. After that, we will show more general numerical  results in order to verify the approximations and extend the calculation beyond this region.

On account of the fast decay of $V(r_\perp, z)$ with the transverse distance $r_\perp$ compared to the much slower decay in the $z$ direction, we include in the sum in eq.~\ref{eq:lambda}  only the ions along the $r_\perp=0$ line (and neglect the fast decaying Yukawa part). That gives
 \bea
 \lambda^{zz}_{zz}= \frac{1}{2\aperp^2\apara} \sum_\Deltab V^{zz}_\Deltab \Delta_z^2 \apara^2
&\approx&
-\frac{Z^2\alpha m_D^2}{2\aperp^2} \sum_{\Delta_z=1}^\infty \Delta_z^2 \frac{d^2}{dz^2} 
 \left(  
  \frac{e^{-z/\lambda_T  }}{4k_e^2+\frac{m_D^2}{2}\log(4k_e z)}
  \frac{ \cos(2 k_e z)}{z}  f\left(  \frac{2k_e^2+m_D^2 \log(4k_e z)/4}{eB}\right)\right)|_{z=\frac{\pi}{ke}\Delta_x}\nn\\
&\approx&
\left\{        
\begin{matrix}
\frac{Z^2\alpha m_D^2}{4\aperp^2\apara^2}\lambda_T^2, 
\qquad \frac{m_D^2}{8k_e^2}\log(4\pi\lambda_T/\apara)\ll 1  \\
\frac{2Z^2\alpha k_e^2}{\aperp^2\apara^2\log(4\pi\lambda_T/\apara)}\lambda_T^2,
\qquad \frac{m_D^2}{8k_e^2}\log(4\pi\lambda_T/\apara)\gg 1  \\
\end{matrix}
\right. \\
&\approx&
\left \{
\begin{matrix}
2.8\times 10^{3}\ MeV^4 \bigpar{     \frac{Z_{28}^4 \rho_8^3}{A_{66}^3B_{15}}   \frac{1}{T_1^2}      }, \\
4.5\times 10^{5}\ MeV^4 \bigpar{     \frac{ Z_{28}^7 \rho_8^6}{A_{66}^6B_{15}^5}   \frac{1}{T_1^2   \log\bigpar{\frac{2.1\times 10^4 Z_{28}^2 \rho_8^2}{A_{66}^2 B_{15}^2 T_1} }  } }. 
\end{matrix}
\right.  \label{eq:lambdazzzz}
\eea

  %%%%%%%%  lambda zzzz   %%%%%%%%%%%%%%%%%%%%%%%%%%%
\bigskip
\begin{figure}[t]\label{fig:lambdazzzz}
  \centerline{\includegraphics{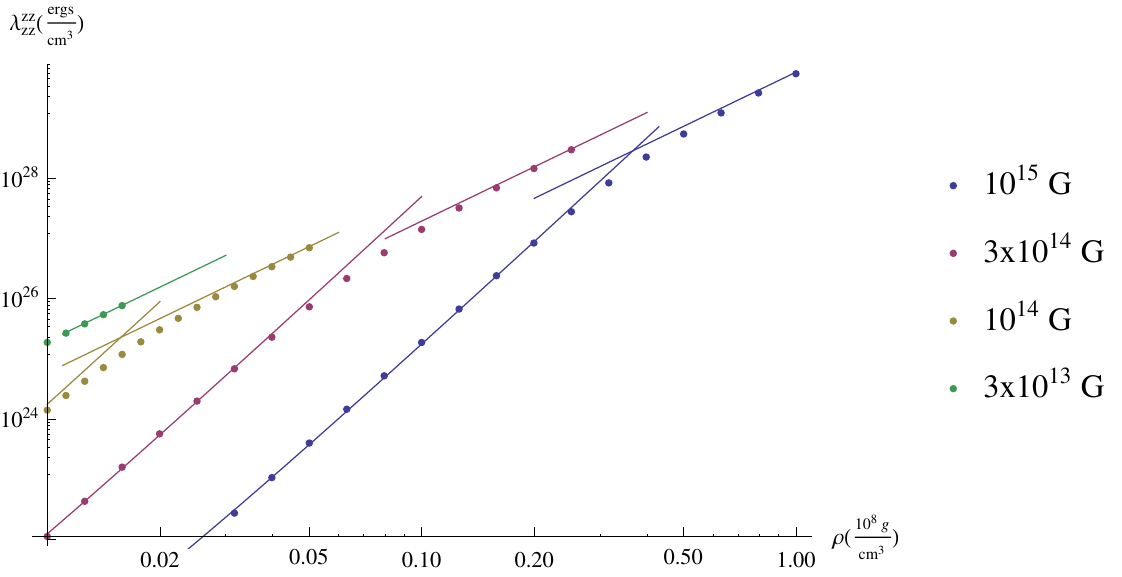}}
%\vskip 0.15in
\noindent
\caption{\label{fig:lambdazzzz} Elastic modulus $\lambda^{zz}_{zz}$ as a function of the density for three different values of the magnetic field and $T=1\ keV$ ($A_{66}=Z_{28}=1$). The lines are the approximate results in eq.~\ref{eq:lambdazzzz}.  We do not show points outside the ``Friedel crystal" zone depicted in fig.~\ref{fig:parameterspace}.
}
\end{figure}  
%%%%%%%%%%%%%%%%%%%%%%%%%%%%%%%%%%%%%%%%%%%%%

Since the sum is nearly divergent (cutoff at $z\sim \lambda_T$ by the exponential), we kept in  eq.~\ref{eq:lambdazzzz} only the most divergent term in the derivative of the potential.  The approximation leading to the upper part of eq.~\ref{eq:lambdazzzz} is valid in the region below the dashed line in fig.~\ref{fig:parameterspace} while the lower part is valid above the same line. 
The quadratic near divergence explains the appearance of the $\lambda_T^2$ factor in the result. Except for the highest temperatures and smallest magnetic fields the contribution in eq.~\ref{eq:lambdazzzz} is  much larger than the electron contribution so
\beq
\lambda^{zz}_{zz}
\approx
5.9\times 10^{29}\frac{ergs}{cm^3} \bigpar{     \frac{Z_{28}^4 \rho_8^3}{A_{66}^3B_{15}}   \frac{1}{T_1^2}      } \label{eq:lambdazzzzanalytic},
\eeq is a convenient analytic expression valid  in most of the parameter space.  This is in sharp contrast to not strongly magnetized crust matter that, like white dwarfs, is held up by the electron pressure. The enhancement comes about due to the fact that each ion  interacts with  thousands of other ions in the same line along the magnetic field.
% %%%%%%%%  Bessel   %%%%%%%%%%%%%%%%%%%%%%%%%%%
%\bigskip
%\begin{figure}[t]\label{fig:bessel}
%  \centerline{\includegraphics{bessel.pdf}}
%%\vskip 0.15in
%\noindent
%\caption{ $K_1\left( a_\perp \sqrt{4 k_e^2 + \frac{m_D^2}{2}\log(4 k_e a_\parallel)} \right)$ as a function of the density $\rho$ for three different values of $B_{15}=1, 0., 0.01$ covering the whole range of parameters we focus on this paper. We used $A_{66}=1$ and $ Z_{28}=1$ . For almost all parameters it is much smaller than one and never exceeds unit, guaranteeing that the $r_\perp\neq 0$ contribution from the Friedel part of the potential are very modest.}
%\label{}
%\end{figure}  
%%%%%%%%%%%%%%%%%%%%%%%%%%%%%%%%%%%%%%%%%%%%%%

The calculation leading to eq.~\ref{eq:lambdazzzz} involves a number of approximations, the most drastic being the neglect of the interactions with ions at $r_\perp \neq 0$. In order to verify the validity of these approximations as well as producing results outside their range of validity we numerically computed the sums in  eq.~\ref{eq:lambda}. Some care is needed in this evaluation. Since the main contribution comes from a sum over   thousand of ions along the line $r_\perp=0$, we should include  a large number of terms in our sum. This can be accomplished by summing a small number and estimating the remaining terms as an integral. The result is shown in fig.~\ref{fig:lambdazzzz} together with the approximate results from eq.~\ref{eq:lambdazzzz}. 
%%%%%%%%%%  lambda xxzz   %%%%%%%%%%%%%%%%%%%%%%%%%%%
\bigskip
\begin{figure}[t]
  \centerline{\includegraphics{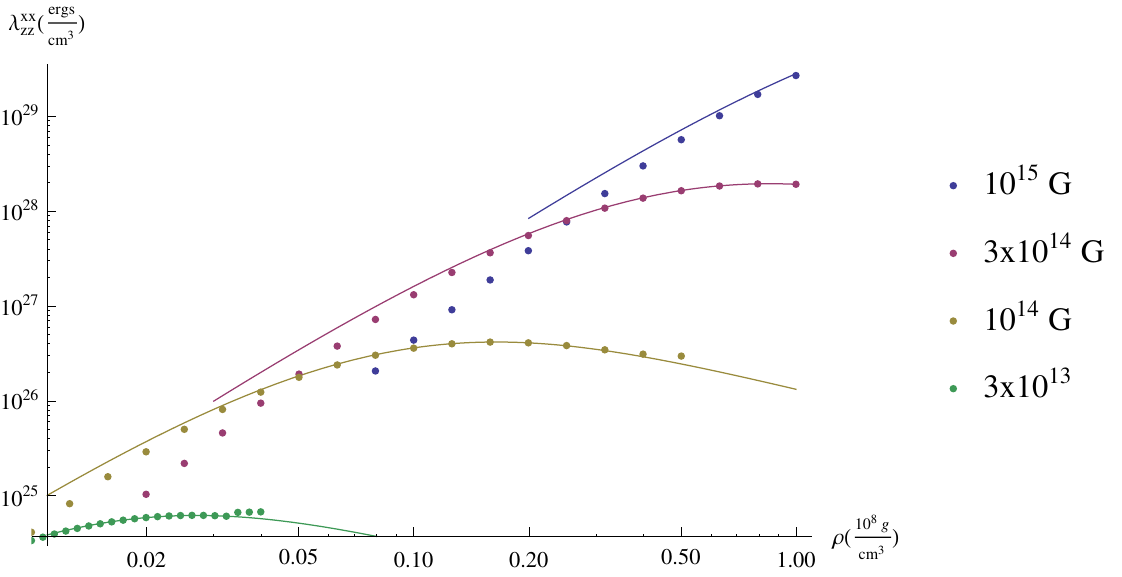}}
%\vskip 0.15in
\noindent
\caption{\label{fig:lambdaxxzz} Elastic modulus $\lambda^{xx}_{zz}$ as a function of the density for four different values of the magnetic field and $T=1\ keV$ ($A_{66}=Z_{28}=1$). The lines are the approximate results in eq.~\ref{eq:lambdaxxzz1}. Only points where we believe $\lambda^{xx}_{zz}$  is dominated by the sum over the $r_\perp=0$ line are shown; for other values the interaction between ions at different $r_\perp$ dominates which can be significantly affected by a different transversal crystal structure. The discrepancy between the points shown and the lines are in the region  $\frac{m_D^2}{8k_e^2}\log(4\pi\lambda_T/\apara)\gg 1 $ where a useful analytic expression was not found but where we still trust the dominance of the $r_\perp=0$ line.
}
\end{figure} 
%%%%%%%%%%%%%%%%%%%%%%%%%%%%%%%%%%%%%%%%%%%%%

Next, we calculate the $\lambda^{xx}_{zz}$ component of the elastic moduli tensor.  We cannot use eq.~\ref{eq:V_ion} for this purpose as the second derivative of the modified Bessel function of the first kind diverges at $r_\perp=0$. This is because our expression of $V_{friedel}$ in eq.~\ref{eq:V_ion} breaks down for $r_{\perp}\alt\frac{1}{\sqrt{eB}}$. To obtain an expression of $V_{friedel}$ that is valid for $r_{\perp}\alt\frac{1}{\sqrt{eB}}$, we begin with an expression in \cite{Sharma:2010bx} that is valid for all $r_{\perp}$ and is given by
\bea
V_{friedel}
&=&-\frac{2\cos(2k_e z)}{\pi z}\int_0^{\infty}dq_{\perp}q_{\perp}J_0(q_{\perp}r_\perp)\frac{\exp(-\frac{q_{\perp}^2}{2eB})e^{-\frac{z}{\lambda_T}}m_D^2(\pi /4)}{(q_{\perp}^2+(2k_e)^2+\exp(-\frac{q_{\perp}^2}{2eB})\frac{m_D^2}{2}\log(4k_e z))^2+\left(\exp(-\frac{q_{\perp}^2}{2eB})\left(\frac{\pi}{4}\right)m_D^2\right)^2}
.
\eea
For $r_{\perp}\agt\frac{1}{\sqrt{eB}}$, the integral is dominated by $q_\perp \alt \sqrt{eB}$ and the exponentials are approximately 1. In this limit, along with the approximation $k_f\gg m_D$ we are lead to our expression for $V_{friedel}$ in eq.~\ref{eq:V_ion}. However, for $r_{\perp}\alt\frac{1}{\sqrt{eB}}$, i.e, $q_\perp \agt \sqrt{eB}$, the exponential in the numerator is crucial for the convergence of the second derivative of $V_{friedel}$. The exponentials in the denominator are negligible in this limit giving,

\bea
V_{friedel} &=&
-\frac{2\cos(2k_e z)}{\pi z}\int_0^{\infty}dq_{\perp}q_{\perp}J_0(q_{\perp}r_\perp)\exp(-\frac{q_{\perp}^2}{2eB})e^{-\frac{z}{\lambda_T}}\frac{m_D^2(\pi /4)}{(q_{\perp}^2+(2k_e)^2)^2}\nn\\
&=&
-\frac{2\cos(2k_e z)}{\pi z}\int_0^{\infty}dq_{\perp}q_{\perp}\left(1-\frac{q_{\perp}^2r_{\perp}^2}{4}+...\right)\exp(-\frac{q_{\perp}^2}{2eB})e^{-\frac{z}{\lambda_T}}\frac{m_D^2(\pi /4)}{(q_{\perp}^2+(2k_e)^2)^2}.
\eea
Following \eqref{eq:lambda}, in the regime where the sum is dominated by the ions at $r_\perp=0$ we have,
\bea
\lambda^{xx}_{zz} &=&
\frac{a_{\parallel}}{2a_{\perp}^2}\sum_{\Delta_z=-\infty} ^{\infty}V^{xx}_{\Delta_z}\Delta_z^2\nn\\
%&=&
%\frac{a_{\parallel}}{a_{\perp}^2}\sum_{\Delta_z=1} ^{\infty}V^{xx}_{\Delta_z}\Delta_z\Delta_z\nn\\
&=&
\left(\frac{a_{\parallel}}{a_{\perp}^2}\sum_{\Delta_z=1} ^{\infty}\frac{\partial^2V_{friedel}}{\partial r_{\perp}^2}\Delta_z^2\right)|_{z=\frac{\Delta_z\pi}{k_e},r_{\perp}=0}\nn \\
%The last line of \eqref{lambdaxxzz1} approximates $V^{xx}_{\Delta_z}$ by $V^{xx}_{friedel}$ as any analytical result that we can trust should not be dominated by the Yukawa part of the potential.
&=&
\frac{a_{\parallel}}{a_{\perp}^2}Z^2\alpha\sum_{\Delta_z=1}^{\infty}\left(\frac{2k_e}{\Delta_z}\int_{0}^{\infty}dq_{\perp}\frac{q_{\perp}^3}{2}e^{\frac{-q_{\perp}^2}{2eB}}\frac{m_D^2/4}{\left(q_{\perp}^2+(2k_e)^2\right)^2}e^{-\frac{\Delta_z\pi}{\lambda_Tk_e}}\Delta_z^2\right)\nn\\
&=&
\frac{a_{\parallel}}{a_{\perp}^2}Z^2\alpha\sum_{\Delta_z=1}^{\infty}\frac{k_e}{\pi}\frac{m_D^2}{2}\Delta_ze^{-\frac{\Delta_z\pi}{\lambda_Tk_e}}\frac{1}{8}\left(-2-e^{\frac{(2k_e)^2}{2eB}}\frac{\left(2eB+(2k_e)^2\right)}{eB}E_i\left(0,\frac{-(2k_e)^2}{2eB}\right)\right)\nn\\
&=&
\frac{Z^2\alpha m_D^2\lambda_T^2}{4a_{\perp}^2a_{\parallel}^2}\left(-\frac{1}{2}-\frac{4k_e^2+2eB}{4eB}e^{\frac{2k_e^2}{eB}}E_i\left(-\frac{2k_e^2}{eB}\right)\right)\nn\\
&=&-\frac{2778.78Z_{28}^4\rho_8^3}{A_{66}^3B_{15}T_1^2}\left(\frac{1}{2}+\frac{1}{B_{15}}0.043e^{\frac{0.15Z_{28}^2\rho_{8}^2}{A_{66}^2B_{15}^3}}\left(11.63B_{15}+\frac{1.7Z_{28}^2\rho_8^2}{A_{66}^2B_{15}^2}\right)Ei\left(\frac{-0.15Z_{28}^2\rho_{8}^2}{A_{66}^2B_{15}^3}\right)\right)\text{MeV}^4\nn\\
&=&-\frac{5.89\times 10^{29}Z_{28}^4\rho_8^3}{A_{66}^3B_{15}T_1^2}\left(\frac{1}{2}+\frac{1}{B_{15}}0.043e^{\frac{0.15Z_{28}^2\rho_{8}^2}{A_{66}^2B_{15}^3}}\left(11.63B_{15}+\frac{1.7Z_{28}^2\rho_8^2}{A_{66}^2B_{15}^2}\right)Ei\left(\frac{-0.15Z_{28}^2\rho_{8}^2}{A_{66}^2B_{15}^3}\right)\right)\text{ergs/cm}^3 \label{eq:lambdaxxzz1}
\eea
where $E_i$ stands for the Exponential Integral function. We were unable to obtain a useful analytical expression in the opposite regime (above the dashed one in fig.~\ref{fig:parameterspace}) where   $\frac{m_D^2}{8k_e^2}\log(4\pi\lambda_T/\apara)\gg 1$. These results are plotted in fig.~\ref{fig:lambdaxxzz}. This figure shows the regions where the approximation used to derive eq.~\ref{eq:lambdaxxzz1} breaks down. The scaling $\lambda^{xx}_{zz}\sim 1/T^2$, predicted by eq.~\ref{eq:lambdaxxzz1}, was also verified numerically. 
 
 The origin of the large value of $\lambda^{xx}_{zz}$ may seem puzzling since the main force acts along the $z$ direction. However, ven though the force keeping the Friedel crystal structure is along the longitudinal direction, it acts on a narrow angle around the $z$ axis. A small deformation of the lattice in the transverse direction misaligns the ion in relation to the $z$ direction and the lattice energy looses the large attractive contribution coming from the $r_\perp$ line. Consequently the energy of the lattice increases quickly and $\lambda^{xx}_{zz}$ can be as large as $\lambda^{zz}_{zz}$.
 It should be noted that our definition of $\lambda^{zz}_{zz}$ and $\lambda^{xx}_{zz}$ is similar to the definition of $b_{11}$ and $c_{44}$ used in some of the previous references on elastic properties of neutron stars crust, such as references \cite{1991ApJ...375..679S,Horowitz:2008xr}. Due to the long-range Friedel potential in the presence of a strong magnetic field our results for these elastic constants in the outer crust of magnetars are more than $10^4$ times larger than the results found in those references for a non-magnetized neutron star crust, and therefore the speed of sound in our case is much larger. In fact in some regions of our parameter space it can get close to the speed of light. It should be noticed that having a speed of sound larger than the speed of light is not physical and can indicate the break-down of some of our approximations. For example when the speed of sound gets large, at some point the static approximation for the ion-ion potential breaks down.

\section{Phonons}
The vibrations of the ion lattice  are responsible for many of the properties of the crust like the specific heat, heat conduction and shear viscosity. In the quantum theory, these vibrations are described by phonons whose dispersion relations are determined by the elastic moduli. The constants $\lambda^{zz}_{zz}$ and $\lambda^{zz}_{xx}$ computed above are sufficient to determine the dispersion relation of phonons moving along the direction of the magnetic field. In this section, we compute these dispersion relations.
 
 We start by considering the action of the ion lattice:
 \beq\label{eq:S}
 S =
 \int_B dt d^3p \left[
 \frac{AMn}{2} |\dot\xib_p|^2 + Ze n \dot\xi_\pb^i \xi^j_{-\pb} \epsilon_{ijk} B^k
 -\frac{n_e \kappa}{2}p_ip_j \xi^i_\pb \xi^j_{-\pb}
 -2 n \sum_\Deltab V^{ij}_\Deltab  \sin^2\left( \frac{\pb.\ab.\Deltab}{2}\right)\xi^i_\pb \xi^j_{-\pb}
 \right],
 \eeq  where $n=\frac{1}{\aperp^2\apara}$ is the number density of ions and $n_e=Zn$ is the number density of electrons. 
 
The first term in eq.~\ref{eq:S} is the kinetic term, the second term describes the interaction between the moving ions and the magnetic field and the remaining ones describe the elastic terms discussed in the previous section. Terms with higher powers of $\xib$ should be added when phonon-phonon interactions are discussed and they can be computed in a manner similar to the computation of the quadratic terms
 
 Let us consider now a phonon moving in the $z$ direction with momentum $\ppara \ll \pi/\apara$ and frequency $\omega$. In this case the action reduces to
 \beq
 S =  \int_B dt d^3p
 \begin{pmatrix}
 \xi_{\pb}^1 &  \xi_{\pb}^2 &  \xi_{\pb}^3
 \end{pmatrix}
 \begin{pmatrix}
   \frac{AMn}{2}\omega^2 - \ppara^2 \lambda^{xx}_{zz}     &      i Zen \omega B  &    0\\
    -i Zen \omega B        &       \frac{AMn}{2}\omega^2 - \ppara^2 \lambda^{yy}_{zz}    &     0             \\
           0    &       0    &          \frac{AMn}{2}\omega^2 -\frac{n_e \kappa}{2}\ppara^2  - \ppara^2 \lambda^{zz}_{zz}\\
 \end{pmatrix}
 \begin{pmatrix}
 \xi_{-\pb}^1 \\
   \xi_{-\pb}^2 \\
     \xi_{-\pb}^3
 \end{pmatrix}.
 \eeq The phonons dispersion relation is given by the condition that the determinant of the matrix above vanishes. We obtain one mode, polarized in the longitudinal direction ($\xib \sim \hat \zb$) with dispersion relation given by(using the fact that $\frac{n_e\kappa}{2} \ll \lambda^{zz}_{zz}$)
 \beq
 \omega = \sqrt{\frac{ 2 \lambda^{zz}_{zz}  }{AMn} }\ \ppara
 \eeq and two transverse modes ($\xib \sim \hat\xb \pm i \hat\yb$), with a polarization rotating about the magnetic field, with dispersion relations given by
 
 \bea
 \omega &=& \sqrt{\frac{Z^2 e^2 B}{A^2M^2} + \frac{2 \lambda^{xx}_{zz}}{AMn}\ppara^2 } \pm \frac{ZeB}{AM}\nn\\
 &\stackrel{\longrightarrow}{\ppara \rightarrow 0}&
 %\new{ \frac{ \lambda^{xx}_{zz}  }{nZeB} \ppara^2 + \cdots}\\
 \left\{
\begin{matrix}
\frac{2ZeB}{AM}+\cdots \approx 5\ keV \bigpar{\frac{ B_{15}   Z_{28} }{A_{66}}  }\\
 \frac{  \lambda^{xx}_{zz}  }{nZeB} \ppara^2 + \cdots
 \end{matrix}
 \right.  ,
 \eea where in the second line we displayed the small $\ppara$ behavior.Phonons are Goldstone bosons arising due to the breaking of translation invariance by the lattice and, as such, are expected to be massless. In  Lorentz invariant theories this implies a linear dispersion relation for the phonons\cite{Goldstone:1962es}, but in the absence of Lorentz invariance, as in the present case, more general dispersion relations are possible. As shown in   \cite{Nielsen:1975hm}, gapless modes with quadratic dispersion relation can appear and count as two linear modes. That is exactly what we find: one quadratic mode and one linear mode corresponding to the three symmetries spontaneously broken (the translations along the three axis). The spectrum composed of a gapped, one quadratic and one linear massless mode was already found in \cite{1980JETP...51..148U}. The speed of sound in the longitudinal direction is given by $ c_s=\sqrt{\frac{ 2 \lambda^{zz}_{zz}  }{\rho} }$. Using eq.~\ref{eq:lambdazzzzanalytic} we can write
\beq
\frac{c_s}{c}\simeq 0.36  \bigpar{     \frac{Z_{28}^2 \rho_8}{A_{66}^{3/2}B_{15}^{1/2}}   \frac{1}{T_{10}}      },
\eeq 
where c is the speed of light and $T_{10}=\frac{T}{10 keV}$.

\section{Conclusion}

In this paper we studied the effect of strong magnetic fields on the structure and elastic properties of neutron star crusts. At low densities and high magnetic fields, similar to the situation that can be found in the outer crust of magnetars, the screening of ion-ion Coulomb potential by electrons is anisotropic and the potential shows a very long-range oscillatory behavior (Friedel oscillations) in the direction along the magnetic field. This long-range potential forces the ions to organize themselves into strongly coupled filaments along the magnetic field, which makes their mechanical properties to be very different from those of not strongly magnetized crusts. Since we haven't computed the exact crystal structure under these conditions, here for simplicity we considered tetragonal lattice and computed the elastic constants that are dominated by the longitudinal structure of the lattice. We found these elastic constants to be significantly larger than that of a bcc Coulomb crystal of comparable densities. In fact, the Friedel crystals discussed in this paper are so hard as to make the speed of sound approach (but not exceed) the speed of light. In this respect, we note that the harmonic approximation will have a limited range of validity in Friedel crystals. This is clear from fig.~\ref{fig:Eion}. The bottom of the potential, near the logarithmic singularity, is very steep but for very modest deformations the energy reaches a much flatter, and consequently much softer, region. For the same reason, even a perfect Friedel crystal with no defects will be very brittle. We leave the quantification of these remarks for  a future publication.

The transitions separating Freidel crystal from other crystal (or liquid) states, shown in fig.~\ref{fig:parameterspace} as the lines separating the dark blue areas from the white ones, are all first order. This implies that the Friedel crystal state can also exist as a metastable state outside the regions shown in fig.~\ref{fig:parameterspace}. This is also evident by the shape of the lattice energy as a function of $\apara$ shown in fig.~\ref{fig:Eion}. 

We would like to stress again that a final word about the true structure of the  lattice can only be determined by a more careful Monte Carlo evaluation. Our arguments in this paper aim at providing a quantitative guide for numerical explorations as well as to bring the attention to the high values of elastic moduli that can be achieved if the Friedel structure of the lattice is confirmed. A full calculation of all elastic moduli would allow us to determine the dispersion relation of phonons moving in all directions. A number of useful properties of the crust would follow from them, for instance the specific heat, heat conductivity (due to umklapp processes) and  shear viscosity. However, these elastic moduli can be reliably computed only after the lattice structure, including the transverse direction, is determined. Assuming a Friedel lattice actually forms (in a range of parameters) with ions separated by $\apara=\pi/k_e$ along the longitudinal direction, two possible lattice structures suggest themselves. One possibility is that the filaments in the $z$ direction avoid each other (due to the Yukawa part of the potential) and form a triangular lattice  in the transverse direction, resulting in a hexagonal lattice. The other is that the nearest-neighbor  filaments shift in the $z$ direction by $\apara/2$ resulting in a body-centered tetragonal lattice. Finally, due to the difference between transverse and longitudinal forces in the system, a state that keeps order in the longitudinal direction but melts in the transverse direction, resulting in a nematic liquid crystal, is a definite possibility. We plan to clarify this issue in a future publication.

\acknowledgments
The authors acknowledge the support of the
U.S. Department of Energy through grant number DEFG02- 93ER-40762.

 %%%%%%%%  eps   %%%%%%%%%%%%%%%%%%%%%%%%%%%
%\bigskip
%\begin{figure}[!htbp]
%  \centerline{{\epsfxsize=3.0in \epsfbox{                  }}}
%%\vskip 0.15in
%\noindent
%\caption{}
%\label{}
%\end{figure}  
%%%%%%%%%%%%%%%%%%%%%%%%%%%%%%%%%%%%%%%%%%%%%

%%%%%%%%%%   PDF and others
%bigskip
%\begin{figure}[!tbp]
%  \centerline{\includegraphics[width=8cm]{coulomb.pdf}}}
%%\vskip 0.15in
%\noindent
%\caption{Examples of graphs of the dominant contribution to $V(\ded,\de)$.}
%\label{fig:coulomb}
%\end{figure}  
%%%%%%%%%%%%%%%%%%

%%%%%%%%%
\bibliographystyle{unsrt}
\bibliography{crust} 
%%%%%%%%%%%%%%%%%%%%%%

\end{document}